\newif\ifcnf
\cnffalse

\ifcnf

\documentclass[times,twocolumn,final]{elsarticle}
\usepackage{cnf}
\usepackage{framed,multirow}
\biboptions{sort&compress}

\else
\documentclass[]{article}
\usepackage[margin=1in,letterpaper]{geometry} 
\usepackage{cite}

\fi

\newif\iffigures
\figurestrue

\newif\ifsubfig
\subfigtrue

\usepackage{booktabs}
\usepackage{amsmath,amsfonts,amssymb}
\usepackage{amsthm}
\usepackage{mathrsfs}
\usepackage[dvipsnames]{xcolor}
\usepackage{bm,relsize}
\usepackage{physics}
\usepackage{graphicx}
\usepackage{blindtext}
\usepackage{pgfplots}
\usepackage{array}
\usepackage{minted}
\usepackage{graphicx}
\usepackage[hidelinks]{hyperref}

\usepackage[capitalize]{cleveref}
\crefname{equation}{}{}
\numberwithin{equation}{section}

\crefname{presumption}{Presumption}{Presumptions}

\usepackage{algorithm,algpseudocode}
\usepackage{bbm}

\usepackage{xargs}
\usepackage[colorinlistoftodos,prependcaption,textsize=tiny]{todonotes}
\newcommandx{\jfmtodo}[2][1=]{\todo[linecolor=magenta,backgroundcolor=magenta!25,bordercolor=magenta,#1]{#2}}

\usetikzlibrary {arrows.meta}
\usetikzlibrary {3d}
\usepackage[dvipsnames]{xcolor}

\pgfplotsset{
	every axis/.append style={
		legend cell align=left,
		legend style={
			draw=none,
		}
	}
}
\tikzset{
	mark options={
		mark size=1pt
	}
}

\definecolor{Set1.1}{HTML}{e41a1c}
\definecolor{Set1.2}{HTML}{377eb8}
\definecolor{Set1.3}{HTML}{4daf4a}
\definecolor{Set1.4}{HTML}{984ea3}
\definecolor{Set1.5}{HTML}{ff7f00}
\definecolor{Set1.6}{HTML}{ff3a1c}
\definecolor{Set1.7}{HTML}{a65628}

\definecolor{Set2.1}{HTML}{66c2a5}
\definecolor{Set2.2}{HTML}{fc8d62}
\definecolor{Set2.3}{HTML}{8da0cb}
\definecolor{Set2.4}{HTML}{e78ac3}
\definecolor{Set2.5}{HTML}{a6d854}
\definecolor{Set2.6}{HTML}{ffd92f}
\definecolor{Set2.7}{HTML}{e5c494}
\definecolor{Set2.8}{HTML}{b3b3b3}

\colorlet{myRed}{OrangeRed}
\colorlet{myGreen}{PineGreen}
\colorlet{myBlue}{RoyalBlue}
\colorlet{myPurple}{Plum}
\colorlet{myTeal}{TealBlue}

\definecolor{UIOrange}{HTML}{FF5F05}
\definecolor{UIBlue}{HTML}{13294B}

\definecolor{UI1}{HTML}{1D58A7} 
\definecolor{UI2}{HTML}{009FD4} 
\definecolor{UI3}{HTML}{FCB316} 
\definecolor{UI4}{HTML}{006230} 
\definecolor{UI5}{HTML}{007E8E} 
\definecolor{UI6}{HTML}{5C0E41} 
\definecolor{UI7}{HTML}{7D3E13} 

\colorlet{dns}{UIBlue}
\colorlet{nmles}{UI2}
\colorlet{apriori}{UI3}
\colorlet{mlles}{UIOrange}
\colorlet{dyn}{UI4!70}

\renewcommand{\bf}[1]{{\mathbf{#1}}}

\newcommand{\rmf}{{\mathrm f}}
\newcommand{\rma}{{\mathrm a}}

\newcommand{\rmg}{{\mathrm g}}

\newcommand{\rmO}{{\mathrm O}}
\newcommand{\rmc}{{\mathrm c}}
\newcommand{\rmr}{{\mathrm r}}

\newcommand{\opt}{{\mathrm{opt}}}

\newcommand{\scA}{{\textsc a}}
\newcommand{\scB}{{\textsc b}}
\newcommand{\scC}{{\textsc c}}
\newcommand{\scL}{{\textsc l}}
\newcommand{\scT}{{\textsc t}}

\newcommand{\bbE}{{\mathbb E}}

\newcommand{\dotomega}{{\dot\omega}}
\newcommand{\vtheta}{{\vec\theta}}

\newcommand{\DNS}{{\textsc{d}}}
\newcommand{\LES}{{\textsc{l}}}

\newcommand{\DelDNS}{{\Delta_\textsc{d}}}

\newcommand{\DelLES}{{\Delta_\textsc{l}}}

\newcommand{\calN}{{\mathcal{N}}}

\newcommand{\Ni}{{N_\mathrm{i}}}
\newcommand{\No}{{N_\mathrm{o}}}
\newcommand{\Nh}{{N_\mathrm{h}}}

\newcommand{\bfx}{\bf x}
\newcommand{\bff}{\bf f}
\newcommand{\btau}{{\bm \tau}}

\newcommand{\filter}{\mathbb F}

\newcommand{\apri}{\textit{a priori}\xspace}

\newcommand{\apost}{\textit{a posteriori}\xspace}
\newcommand{\etal}{\textit{et al.}\xspace}

\renewcommand{\Re}{\mathrm{Re}}
\newcommand{\Ma}{\mathrm{Ma}}
\renewcommand{\Pr}{\mathrm{Pr}}
\newcommand{\Sc}{\mathrm{Sc}}
\newcommand{\Ka}{\mathrm{Ka}}
\newcommand{\Da}{\mathrm{Da}}

\begin{document}

\title{A physics-constrained machine-learning sub-grid-scale modeling approach for turbulent premixed flames}

\ifcnf

\verso{Suh et al.}

\begin{frontmatter}

\author[1]{Seung Won \snm{Suh}\corref{cor1}}
\author[2]{Jonathan F. \snm{MacArt}}
\author[1]{Luke N. \snm{Olson}}
\author[1]{Jonathan B. \snm{Freund}}

\cortext[cor1]{Corresponding author: University of Illinois Urbana--Champaign, Urbana, Illinois, 61801, United States of America.}
\emailauthor{suh29@illinois.edu}{Seung Won Suh}

\address[1]{University of Illinois Urbana--Champaign, Urbana, Illinois, 61801, United States of America}
\address[2]{University of Notre Dame, Notre Dame, Indiana, 46556, United States of America}

\else

\author{
	Seung Won Suh\thanks{Mechanical Science and Engineering, University of Illinois Urbana--Champaign, United States of America, suh29@illinois.edu},
	Jonathan F. MacArt\thanks{Aerospace and Mechanical Engineering, University of Notre Dame, United States of America, jmacart@nd.edu},
	Luke N. Olson\thanks{Computer Science, University of Illinois Urbana--Champaign, United States of America, lukeo@illinois.edu},
	and Jonathan B. Freund\thanks{Aerospace Engineering, University of Illinois Urbana--Champaign, United States of America, jbfreund@illinois.edu}
}
\maketitle

\fi

\begin{abstract}
	A physics-embedded training framework is used to close the sub-grid-scale dynamics of turbulent premixed flames.
	The trained model augments the resolved flow equations and is trained to match its predicted flow field to trusted data.
	An end-to-end optimization of the coupled resolved equations and embedded model leads to a model that is robust and effective for a freely propagating premixed flame in turbulence, modeled by a single-species, single-step, and irreversible chemical reaction.
	Important constraints are built into the training formulation for conservation, scalar boundedness, and equivariance.
	The model is scaled based on the residual of the resolved flow equations to focus its influence where the closure is needed.
	It outperforms other cases considered~---~no-model, dynamic closure, and the same machine learning model trained directly to fit the residual data~---~for a long-time simulation, correcting the turbulence dissipation and flame kinematics.
	Finally, it is shown how the prediction-based training better informs the model than the precise residual data.
\end{abstract}

\ifcnf

\end{frontmatter}

\linenumbers
\fi

\section{Introduction}
Large-eddy simulation (LES) benefits from explicit closure, especially when predictions rely heavily on the truncated scales~\cite{langford1999optimal}.
Established sub-grid-scale (SGS) models based on relatively simple parameterizations are successful for inert Navier--Stokes turbulence~\cite{piomelli2014large}, yet it is understood that the challenge increases with increasing complexity of the SGS physics, as in the case of the chemical reactions we consider~\cite{steinberg2021structure,pitsch2006large}.
In such cases, a greater model complexity should enable more effective representation of the SGS multiphysics.
Machine learning (ML), as a nonlinear regression tool with many parameters, is therefore attractive if it can be reliably trained~\cite{brenner2019perspective,duraisamy2019turbulence}.

Any ML closure can `fit' data, but that objective does not necessarily reflect the prediction target.
For LES, this is related to the well-known inconsistency between \apri and \apost evaluation of SGS models~\cite{moser2021statistical}.
Effective models do not necessarily match their true counterparts (\apri), while models that show good \apri agreement often fail when applied (\apost)~\cite{bardina1980improved}.
In short, a model that replicates known SGS behavior might fail when used, and a successful model might only loosely track the actual SGS physics.
This is because the discrepancy in the governing equations does not directly represent the discrepancy in the predictions.
Thus, the remarkable fitting capability of ML methods does not guarantee robust performance~\cite{duraisamy2021perspectives}.

Sirignano \etal~\cite{sirignano2020dpm} show this for inert turbulence.
Their ``deep-learning PDE model'' (DPM) trained its embedded ML closure to ensure that the ML-augmented PDE solution matched the target.
So trained, it then successfully predicted the decay of isotropic turbulence, whereas a corresponding \apri-fitted closure was unstable.
The robustness is thought to come from the tight coupling of the training to the PDE, in this case the filtered Navier--Stokes equations.
Since the objective entails a PDE solution with the model, the trained closure is, in a sense, aware of the time integration and predicted flow evolution.
A challenge then is that the end-to-end optimization with gradient descent necessitates an adjoint PDE solve to provide a sensitivity gradient.
Crafted this way, training accounts for both closure and consequences of numerical error, which is another anticipated advantage over \apri-trained models.
An additional benefit of DPM is its flexibility in defining the training objective, which makes it viable for situations with limited data: it does not need SGS data to train an SGS model~\cite{macart2021embedded}.

Applications of DPM with simpler physics include LES of decaying isotropic turbulence~\cite{sirignano2020dpm}, turbulent jets~\cite{macart2021embedded}, flows around bluff bodies~\cite{sirignano2023deep}, and two-dimensional turbulence with particles~\cite{rivera2026training}; RANS simulation for turbulent premixed jet flames~\cite{kakka2025neural}; and active flow control~\cite{liu2024adjoint,liu2025active}.
Other strategies for \apost ML SGS modeling also suggest outperformance of \apri-trained models~\cite{holland2019field,um2020solver,kochkov2021machine,list2022learned,agdestein2025discretize}.
There are also reported successes of \apri ML SGS models for simple flows~\cite{wang2018investigations,beck2019deep,park2021toward,kang2023neural,lozano2023machine,benjamin2024neural}, though their utility for more complex problems is yet to be evaluated.

DPM for LES of chemically reacting turbulence necessitates additional developments.
We demonstrate it for a statistically planar premixed flame in high-speed turbulence with simple chemistry~\cite{suh2025tvd}, although the formulation is not restricted to simple chemistry.
This is a step beyond existing ML SGS closure approaches for reacting turbulence that use \apri training~\cite{ihme2022combustion,lapeyre2019training,xing2022deep,piu2025data}.
A recent effort applied DPM to LES of turbulent premixed jet flames~\cite{kakka2026solver}, though their model modified the scalar eddy viscosity.
We introduce a more comprehensive form for the SGS closure for reacting turbulence, not limited to the eddy-viscosity types, with certain physical constraints directly enforced.

\Cref{sec:simulation} introduces the flow configuration and behavior with DNS data that are also used in the training.
\Cref{sec:closure} introduces constraints on the closure that preserve key physical properties in the model.
\Cref{sec:optimization} discusses training strategies specific to our demonstration, with results in~\cref{sec:les}.
Additional discussion in~\cref{sec:discussion} further illuminates the benefit of the embedded training approach.
Finally,~\cref{sec:conclusion} summarizes conclusions.

\section{Reacting flow simulation} \label{sec:simulation}
\subsection{Governing equations and flow parameters}
\begin{figure}[!t]
	\iffigures
	\centering
    \ifcnf
    \includegraphics[width=0.7\linewidth]{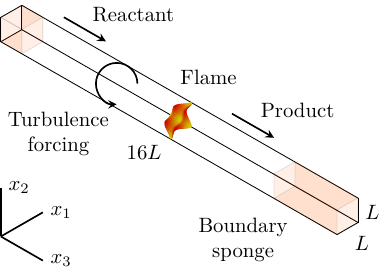}
    \else
    \includegraphics[width=0.4\linewidth]{figures/domain.pdf}
    \fi
	\fi
	\caption{
        Computational domain.
    }
	\label{fig:domain}
\end{figure}
A statistically planar premixed flame propagates through forced isotropic turbulence (\cref{fig:domain}).
The governing equations are the fully compressible Navier--Stokes equations with a single-step irreversible chemical reaction:
\begin{equation} \label{eqn:goveq-continuous}
	\pdv{\vec q}{t} = -\nabla \cdot \vec{\bf f\,}(\vec q) + \vec s(\vec q) \equiv \vec R(\vec q)
\end{equation}
for
\begin{equation}
	\vec q(\bfx, t) = \big[\, \rho \;\; \rho \bf u \;\; \rho E \;\; \rho Y \, \big]^\top.
\end{equation}
The corresponding right-hand side is
\begin{equation}
	\vec R = \big[\, R_\rho \;\; (R_{\rho u_1}, R_{\rho u_2}, R_{\rho u_3})^\top \;\; R_{\rho E} \;\; R_{\rho Y} \, \big]^\top,
\end{equation}
with coordinates $\bf x = (x_1, x_2, x_3)^\top$ parametrizing space, $\rho$ the density, and $\bf u = (u_1, u_2, u_3)^\top$ the velocity.
The total specific sensible energy $E$ is
\begin{equation} \label{eqn:total-energy}
	E = \frac{T}{\gamma} + \frac{\bf u \cdot \bf u}{2},
\end{equation}
where $T$ is the temperature, and $\gamma$ is the specific heat ratio.
Pressure $p$ is given by the ideal gas equation of state,
\begin{equation}
	p = \frac{\gamma-1}{\gamma} \rho T,
\end{equation}
and $Y$ is the reactant mass fraction.
Physical variables are nondimensionalized in the usual way by the unburnt gas density $\rho_0^*$, a reference length $L^*$, a reference velocity $U^*$, and the specific heat $c_p^*$:
\begin{equation}
    \begin{gathered}
	    \rho = \frac{\rho^*}{\rho^*_0}, \quad \bf x = \frac{\bf x^*}{L^*}, \quad \bf u = \frac{\bf u^*}{U^*}, \quad t = \frac{U^* t^*}{L^*}, \\
        p = \frac{p^*}{\rho^*_0 U^{* 2}}, \quad \text{and} \quad T = \frac{c_p^* T^*}{U^{* 2}},
    \end{gathered}
\end{equation}
where the superscript $*$ denotes dimensional quantities.

The flux $\vec{\bf f\,}$ in~\cref{eqn:goveq-continuous} includes inviscid and viscous contributions:
\begin{equation} \label{eqn:flux}
	\vec{\bf f\,}(\vec q) = \begin{bmatrix}
		\rho \bf u \\ \rho \bf u \bf u^\top + p \bf I \\ \bf u(\rho E + p) \\ \rho \bf u Y
	\end{bmatrix} - \begin{bmatrix}
		\bf 0 \\ \bm \tau \\ \bm \tau \cdot \bf u - \bm \varphi_T \\ -\bm \varphi_Y
	\end{bmatrix},
\end{equation}
where
\begin{subequations} \label{eqn:flame-flux-nondim}
	\begin{gather}
		\bm \tau = \frac{\mu}{\Re} \left[ \nabla \bf u + (\nabla \bf u)^\top - \frac{2}{3} (\nabla \cdot \bf u) \bf I \right], \\
		\bm \varphi_T = -\frac{\mu}{\Re\Pr} \nabla T, \quad \text{and} \\
		\bm \varphi_Y = -\frac{\mu}{\Re\Sc} \nabla Y,
	\end{gather}
\end{subequations}
with the Reynolds number, Prandtl number, and Schmidt number
\begin{equation} \label{eqn:RePrSc}
	\Re \equiv \frac{\rho_0^* U^* L^*}{\mu_0^*}, \quad \mathrm{Pr} \equiv \frac{\mu_0^* c_p^*}{k_0^*}, \quad \text{and} \quad \mathrm{Sc} \equiv \frac{\mu_0^*}{\rho_0^* \mathcal D_0^*}.
\end{equation}
The temperature-dependent transport coefficients follow a power law,
\begin{equation}
	\frac{\mu^*}{\mu^*_0} = \frac{k^*/c_p^*}{k^*_0/c_{p,0}^*} = \frac{\rho^* \mathcal D^*}{\rho^*_0 \mathcal D^*_0} = \left( \frac{T^*}{T^*_0} \right)^{0.7},
\end{equation}
where $\mu^*$ is the dynamic viscosity, $k^*$ is the thermal conductivity, and $\mathcal D^*$ is the mass diffusivity.
Finally, the chemical reaction source term~\cref{eqn:goveq-continuous} is
\begin{equation}
	\vec s(\vec q) = \begin{bmatrix}
		0 & \bf 0 & \alpha T_\rmf \dotomega & -\dotomega
	\end{bmatrix}^\top,
\end{equation}
with the adiabatic flame temperature $T_\rmf$ and the heat release ratio
\begin{equation} \label{eqn:heat-release}
	\alpha \equiv \frac{T_\rmf^* - T_0^*}{T_\rmf^*}.
\end{equation}
The Arrhenius reaction rate $\dotomega$ is
\begin{equation} \label{eqn:arrhenius}
	\dotomega = A \rho Y \exp\left[ \frac{\beta}{\alpha} \left( 1 - \frac{T_\rmf}{T} \right) \right],
\end{equation}
where $A$ is nondimensionalized as
\begin{equation} \label{eqn:arrhenius-constant}
	A \equiv A^* \rho_0^* \exp(-\frac{T_\rma^*}{T_\rmf^*}),
\end{equation}
with $T_\rma^*$ the activation temperature, which is also nondimensionalized as the Zel'dovich number
\begin{equation} \label{eqn:zeldovich}
	\beta \equiv \frac{\alpha T_\rma^*}{T_\rmf^*}.
\end{equation}

Dimensionless parameter values in~\cref{table:dimensionless} define the flow conditions.
Most values are adapted from Poludnenko and Oran~\cite{poludnenko2010interaction}, including $\Pr$, $\Sc$, $\alpha$, $\beta$, and $\gamma$, based on a simplified reaction model for a stoichiometric hydrogen--air mixture~\cite{gamezo2008flame}.
The reference Mach number,
\begin{equation} \label{eqn:Ma}
	\Ma \equiv \frac{U^*}{\sqrt{(\gamma-1) c_p^* T_0^*}},
\end{equation}
is lowered here to prevent unwanted transition to detonation~\cite{poludnenko2011spontaneous}.
The reference Reynolds number and the Arrhenius constant are adjusted to intensify the SGS modeling challenge.

\begin{table}
	\centering
	\begin{tabular}{c c c}
		\toprule
        Reynolds number~\cref{eqn:RePrSc} & $\Re$ & $10^4$ \\
		Prandtl number~\cref{eqn:RePrSc} & $\Pr$ & 0.1 \\
		Schmidt number~\cref{eqn:RePrSc} & $\Sc$ & 0.1 \\
        Arrhenius constant~\cref{eqn:arrhenius-constant} & $A$ & 1000 \\
        Heat release ratio~\cref{eqn:heat-release} & $\alpha$ & 0.863 \\
        Zel'dovich number~\cref{eqn:zeldovich} & $\beta$ & 5.49 \\
        Heat capacity ratio~\cref{eqn:total-energy} & $\gamma$ & 1.17 \\
        Mach number~\cref{eqn:Ma} & $\Ma$ & 0.05 \\
		\bottomrule
	\end{tabular}
	\caption{
        Dimensionless parameters.
    }
	\label{table:dimensionless}
\end{table}

\subsection{Numerical methods} \label{sec:numerics}
The computational domain shown in~\cref{fig:domain} has dimensions $(L_1,L_2,L_3) = (1,1,16)$ and is spanwise ($x_1$, $x_2$) periodic.
Absorbing layers at the streamwise ($x_3$) boundaries create nominal far-field conditions and are sufficiently long in $x_3$ to attenuate effects from the domain edge~\cite{freund1997proposed,suh2026thesis}.

State variables are represented on a uniform staggered Cartesian mesh~\cite{harlow1965numerical,nagarajan2003robust,boersma2005staggered} with spacings $\DelDNS$ for DNS and $\DelLES$ for LES.
Second-order central finite differences provide spatial derivatives and linearly interpolate the staggered state variables.
Scalars $T$ and $Y$ are upwinded based on the momentum $\rho \bf u$ to ensure boundedness~\cite{larrouturou1991preserve} with the superbee limiter~\cite{sweby1984high}.

Setting $\DelDNS=1/128$ resolves all relevant scales (reported in more detail elsewhere~\cite{suh2026thesis}), while we will show that $\DelLES=1/16$ omits scales that are important.
Time is uniformly discretized with the time step size $\Delta t_\DNS$ for DNS and $\Delta t_\LES$ for LES and a standard fourth-order Runge--Kutta (RK4) algorithm.
We use $\Delta t_\DNS = 10^{-4}$ for DNS and $\Delta t_\LES = 5\Delta t_\DNS$ for LES.
Simulations of one-dimensional flames, reported in full elsewhere~\cite{suh2026thesis}, show effective convergence for $\Delta \lesssim 2\DelDNS$.
With $\DelLES = 8\DelDNS$, mesh spacing is comparable to the nominal flame thickness ($\ell_\rmf \equiv 1/\max \Vert \nabla Y \Vert_2 = 1.14\DelLES$), so such simulations will be mesh-dependent and will benefit from sub-grid-scale models.
Pre-flame inert turbulence simulations show no change for $\DelDNS \to 0.5\DelDNS$ refinement.
Full discrepancies between the flame LES and DNS are presented in~\cref{sec:challenge}.

\subsection{Simulation results} \label{sec:challenge}
\subsubsection{DNS} \label{sec:dns}
The DNS field $\vec q$ is initialized at time $t=-2$ by superimposing a laminar flame profile for $\rho$, $u_3$, $T$, and $Y$ atop an established inert forced turbulence field.
Transients decay over $t \in [-2,0]$.
Fields are then saved every $100\Delta t_\DNS = 0.01$ for $t \in [0,4]$, producing 401 training instances (see~\cref{sec:embedded}).

The upstream turbulence Reynolds number is
\begin{equation} \label{eqn:turbRe}
	\Re_\mathrm{t} = \frac{u' \ell}{\nu_0} = 376,
\end{equation}
and the turbulence Mach number is
\begin{equation}
	\mathrm{Ma}_\mathrm{t} = \frac{u'}{\sqrt{(\gamma-1)T_0}} = 9.4 \times 10^{-3},
\end{equation}
based on the upstream flow conditions, with root-mean-square velocity fluctuation $u'=0.188$, integral length scale $\ell=0.2$, and kinematic viscosity $\nu_0 = 10^{-4}$.
With the laminar flame speed $s_\scL = 0.169$ and the laminar flame thickness $\ell_\rmf = 0.032$ from the one-dimensional flame simulation, the Damk\"ohler number is
\begin{equation}
	\Da = \frac{\tau_\ell}{\tau_\mathrm{chem}} = \frac{\ell/u'}{\ell_\rmf/s_\scL} = 5.62,
\end{equation}
with $\tau_\ell$ the nominal large-eddy turnover time and $\tau_\mathrm{chem}$ the chemical time scale.
The Karlovitz number is
\begin{equation}
	\Ka = \frac{\tau_\mathrm{chem}}{\tau_\eta} = \frac{\ell_\rmf/s_\scL}{(\nu_0/\varepsilon_\mathrm{in})^{1/2}} = 3.09,
\end{equation}
with $\tau_\eta$ the Kolmogorov time scale, and $\varepsilon_\mathrm{in}$ is the energy injection rate by the turbulence forcing~\cite{eswaran1988examination}, which balances the turbulence dissipation rate for our statistically stationary flow.

\begin{figure}
	\centering
	\iffigures
    \ifcnf
	\includegraphics[width=\linewidth]{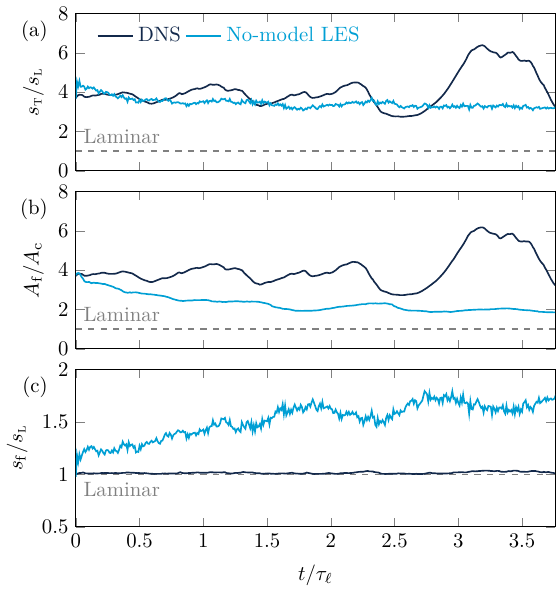}
    \else
	\includegraphics[width=0.6\linewidth]{figures/results_flame_challenge.pdf}
    \fi
	\fi
    \caption{
        Flame-related predictions: (a) mean turbulent flame speed $s_\scT$~\cref{eqn:st}, (b) surface area $A_\rmf$ of the isosurface based on $Y=0.18$, and (c) average local flame speed $s_\rmf$~\cref{eqn:sf} normalized by laminar flame values (dashed lines).
    }
	\label{fig:challenge-flame}
\end{figure}

The primary quantities of interest (QoIs) are the turbulent flame speed
\begin{equation} \label{eqn:st}
	s_\scT \equiv \sum_{\bfx \in \mathrm{mesh}} \dotomega(\bfx) \Delta V,
\end{equation}
where $\Delta V$ is the cell volume, the flame surface area $A_\rmf$, and the average local flame speed
\begin{equation} \label{eqn:sf}
	s_\rmf \equiv \frac{s_\scT A_\rmc}{A_\rmf},
\end{equation}
with $A_\rmc = L_1 L_2$.
The flame surface area $A_\rmf$ is calculated based on the isosurface of $Y=\mathrm{argmax}_Y \dotomega = 0.18$, which is computed via the marching cube algorithm~\cite{lewiner2003efficient} as supported by the \texttt{scikit-learn} package in \texttt{Python}~\cite{van2014scikit}.
DNS predictions are shown in~\cref{fig:challenge-flame} with a visualization in~\cref{fig:flame3d}(a).
The large fluctuations of $s_\scT$ and $A_\rmf$ and their high correlation in~\cref{fig:challenge-flame}(a) and~\cref{fig:challenge-flame}(b) are the consequences of large eddies folding the thin flame, as expected for $u'/s_\scL = 1.11$~\cite{fogla2015effect}.
Flame folding is a dominant factor in our QoIs that will need to be accurately represented.

\subsubsection{LES without explicit closure (no-model LES)} \label{sec:nles}
The no-model LES field $\vec Q$ is initialized as the filtered DNS field $\filter \vec q$ at $t=0$ and solves~\cref{eqn:goveq-continuous} without an explicit closure.
With fixed downsampling $\DelLES/\DelDNS=8$, the truncating operator $\filter$ first applies an explicit box filter to $\vec q$ with width $\bar\Delta = 2\DelDNS$ to reduce the aliasing from the following downsampling.

Kinetic energy spectra (which will be examined in~\cref{fig:results-turb+flame}(a)) show elevated small-scale turbulence energy due to lack of dissipation, as expected for a non-dissipative numerical scheme.
\Cref{fig:challenge-flame} shows the modeling challenge through comparison of the DNS and the no-model LES for $s_\scT$, $A_\rmf$, and $s_\rmf$.
The apparent good match in the mean $s_\scT$ actually masks counteracting errors in $A_\rmf$ and $s_\rmf$.
The root problem is apparently the failure of the no-model LES to represent the flame folding.
(The instance visualized in~\cref{fig:flame3d}(b) is typical, with a nearly flat flame across most of the span.)
The enhanced $s_\rmf$ over $s_\scL$ in~\cref{fig:challenge-flame}(c) also implies missing SGS turbulence--flame interactions in the no-model LES.
The LES overly excites upstream turbulence while also increasing $s_\rmf$ and suppressing flame cusps, leading to a stable flat flame.

\section{Closure model} \label{sec:closure}
\subsection{Basic formulation}
A trainable closure model $\vec M$ is introduced into~\cref{eqn:goveq-continuous} as
\begin{equation} \label{eqn:goveq-les-ml}
	\pdv{\vec Q^\theta}{t} = \vec R(\vec Q^\theta) + \vec M(\vec Q^\theta; \vtheta) \equiv \vec R^\theta,
\end{equation}
with parameters $\vtheta$.
We train $\vec M$ to fit $\vec Q^\theta$ to the trusted $\filter \vec q$ as closely as possible.

In designing the specific $\vec M$, a reduced model, some properties can and should be exactly preserved if possible.
Constraints can be learned, of course, but only approximately.
As such, it is beneficial to build them into the model.
For example, for hyperbolic systems, unconstrained closures cannot perfectly learn the advantageous TVD property~\cite{suh2025tvd}.
This will be generalized here to incorporate additional constraints.

At the outset, it is not clear which constraints to enforce.
There are infinitely many moments that could be cast as mathematical constraints for a fluctuating turbulence field.
Whether a property is deemed essential depends on the QoI: some constraints must be satisfied exactly for either qualitative or quantitative correctness, while others may be violated as long as they do not overly degrade the fidelity of the simulation, such as the divergence-free constraint~\cite{sirignano2020dpm} or the symmetry of the SGS stress tensor~\cite{macart2021embedded}.
In the following subsections, we start with an unconstrained basic model, then introduce constraints.

\subsection{NN architecture} \label{sec:nn}
A fully-connected architecture with gating layers is used:
\begin{equation} \label{eqn:mlp}
	\begin{split}
		z^1 &= \tanh(W^0 y + b^0) \\
		z^2 &= \tanh(W^1 z^1 + b^1) \\
		z^3 &= z^2 \odot z_\rmg^1 \quad \text{with} \quad z_\rmg^1 = \tanh(W^2 y + b^2) \\
		z^4 &= \tanh(W^3 z^3 + b^3) \\
		z^5 &= z^4 \odot z_\rmg^2 \quad \text{with} \quad z_\rmg^2 = \tanh(W^4 y + b^4) \\
		\phi^\calN (y) &= W^5 z^5 + b^5,
	\end{split}
\end{equation}
as introduced by Sirignano and Spiliopoulos~\cite{sirignano2018dgm} and used subsequently~\cite{sirignano2020dpm,macart2021embedded,sirignano2023deep,nair2023deep,liu2024adjoint,rivera2026training}.
Denoting $\Ni$, $\No$, and $\Nh$ as the number of units in the input layer, the output layer, and the hidden layer, respectively, the weights and biases have the following dimensions:
\begin{equation}
	\begin{split}
		W^0, W^2, W^4 &\in \mathbb R^{\Nh \times \Ni}, \\
		W^1, W^3 &\in \mathbb R^{\Nh \times \Nh}, \\
		W^5 &\in \mathbb R^{\No \times \Nh}, \\
		b^0, b^1, b^2, b^3, b^4 &\in \mathbb R^\Nh, \quad \text{and} \\
		b^5 &\in \mathbb R^\No.
	\end{split}
\end{equation}
For all models used, $\Nh=50$ and $\Ni=40$ based on local $\vec Q^\theta$ around cell centers:
\begin{equation} \label{eqn:input}
	\big[ \bf D \bf u, c, \bf D^2 [\bf D \bf u, c]^\top \big]^\top \in \mathbb R^\Ni, 
\end{equation}
where $c \equiv 1-Y$ is the flame progress variable, and
\begin{equation}
    \begin{gathered}
        \bf D \phi(\bfx) = \begin{bmatrix}
            \phi(\bfx + \frac{\Delta}{2}\bf e_1) - \phi(\bfx - \frac{\Delta}{2}\bf e_1) \\
            \phi(\bfx + \frac{\Delta}{2}\bf e_2) - \phi(\bfx - \frac{\Delta}{2}\bf e_2) \\
            \phi(\bfx + \frac{\Delta}{2}\bf e_3) - \phi(\bfx - \frac{\Delta}{2}\bf e_3)
        \end{bmatrix} \quad \text{and} \\
        \bf D^2 \phi(\bfx) = \begin{bmatrix}
        \phi(\bfx + \Delta \bf e_1) -2\phi(\bfx) + \phi(\bfx - \Delta \bf e_1) \\
        \phi(\bfx + \Delta \bf e_2) -2\phi(\bfx) + \phi(\bfx - \Delta \bf e_2) \\
        \phi(\bfx + \Delta \bf e_3) -2\phi(\bfx) + \phi(\bfx - \Delta \bf e_3)
        \end{bmatrix}
    \end{gathered}
\end{equation}
compute the differences of a state $\phi$ between nearest mesh points, with $\bf e_i$ the unit vector in the $i$-th spatial dimension.
Finally, any quantity with the superscript $\mathcal N$ refers to this elementary model~\cref{eqn:mlp}.

\subsection{Constrained ML representation}
\begin{figure}
    \centering
    \ifcnf
    \includegraphics[width=0.8\linewidth]{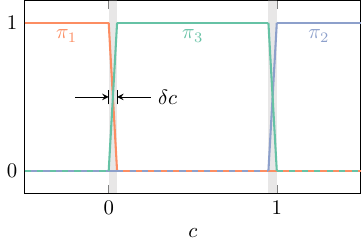}
    \else
    \includegraphics[width=0.4\linewidth]{figures/mixingweights.pdf}
    \fi
    \caption{
        Weights $\pi_k$~\cref{eqn:mixture} for model localization versus $c$.
    }
    \label{fig:mix-weight}
\end{figure}

The full model is introduced here and explained through this section: 
\begin{equation} \label{eqn:closure}
	\vec M(\vec Q^\theta; \vtheta) = \underbrace{\begin{bmatrix}
		0 \\ \nabla \cdot \bm \tau^\rmc \\ \nabla \cdot (\bm \tau^\rmc \bf u) \\ 0
	\end{bmatrix}}_{\vec M^\scA} - \underbrace{\begin{bmatrix}
		0 \\ \bf 0 \\ \nabla \cdot (\bff^\rmc T) \\ \nabla \cdot (\bff^\rmc Y)
	\end{bmatrix}}_{\vec M^\scB} + \underbrace{\begin{bmatrix}
		0 \\ \bf 0 \\ \alpha T_\rmf s^\rmc \\ -s^\rmc
	\end{bmatrix}}_{\vec M^\scC}.
\end{equation}
Because $\btau^\rmc$ is symmetric, its independent components are modeled as a mixture of elementary NNs $\vec g^{\calN(k)} (\vec Q^\theta; \vtheta_g^{\,(k)}) = [g^{(k)}_1, \cdots, g^{(k)}_6]^\top$:
\begin{equation} \label{eqn:closure-turb}
    [\tau^\rmc_{11}, \tau^\rmc_{22}, \tau^\rmc_{33}, \tau^\rmc_{12}, \tau^\rmc_{23}, \tau^\rmc_{13}]^\top = C_\tau \sum_{k=1}^3 \pi_k \langle \vec g^{\calN(k)} \rangle_G,
\end{equation}
where each $\vec g^{\calN(k)}$ has $\No=6$ output units for $k \in \{1,2,3\}$.
The operation $\langle \, \rangle_G$ enforces the frame-invariance, as defined in~\cref{sec:equivariance}.
The constant $C_\tau$ rescales the output of NN $\vec g^{\calN(k)}$ for the residual-consistency as discussed in~\cref{sec:apricon}.
The weight $\pi_k$ assigns each $\vec g^{\calN(k)}$ to three distinct regimes as a switching function with linear blending:
\begin{equation} \label{eqn:mixture}
	\pi_k(\vec Q^\theta) = \begin{cases}
		\min \{ \max \{ 1-\frac{c}{\delta c}, 0 \}, 1\} & k=1 \text{ (unburnt)} \\
		\min \{ \max \{ 1+\frac{(c-1)}{\delta c}, 0 \}, 1\} & k=2 \text{ (burnt)} \\
		1-\pi_1-\pi_2 & k=3 \text{ (flame),}
	\end{cases}
\end{equation}
where $\delta c = 0.05$ is the width of the transition between regimes, as illustrated in~\cref{fig:mix-weight}.

For scalar $T$ and $Y$ transport, the flux $\bff^\rmc$ and the source $s^\rmc$ are built upon another elementary NN model $\vec h^\calN(\vec Q^\theta; \vtheta_h) = [h_1, h_2, h_3, h_4]^\top$ with $\No=4$:
\begin{equation} \label{eqn:closure-chem}
	\bff^\rmc = \nabla \times \langle [h_1, h_2, h_3]^\top \rangle_G \quad \text{and} \quad	s^\rmc = C_s \langle h_4 \rangle_G.
\end{equation}
The prefactor $C_s = \dotomega$ rescales the output of $h_4$ as $C_\tau$ does for the residual-consistency discussed in~\cref{sec:apricon}.
Altogether, the NN model is $\vec\calN \equiv [ \{\vec g^{\calN(k)}\}_{k=1}^3; \; \vec h^\calN ]$ with $N_\theta = 46{,}122$ parameters $\vtheta \equiv [ \{ \vtheta_g^{(k)} \}_{k=1}^3; \; \vtheta_h ]$.

\subsubsection{Conservation} \label{sec:conservation}
Casting $\vec M^\scA$ and $\vec M^\scB$~\cref{eqn:closure} in a divergence form, the model creates no bulk acceleration.
Additional conservation laws follow in~\cref{eqn:closure}, including total energy conservation.
Using $\nabla \cdot (\btau^\rmc \bf u)$ in $\vec M^\scA$ ensures $\btau^\rmc$ only affects the kinetic energy, not the internal energy.
The last two components of $\vec M^\scC$ in~\cref{eqn:closure}, $\alpha T_\rmf s^\rmc$ and $-s^\rmc$, are coupled so that the total energy is conserved and $\vec M^\scC$ is adiabatic: the chemical energy destroyed in the $\rho Y$-equation by the model $s^\rmc$ is identical to the mechanical energy created in the $\rho E$-equation, $\alpha T_\rmf s^\rmc$.
Finally, with $\bm \tau^\rmc$ in $\vec M^\scA$ symmetric, the model does not add angular momentum.

\subsubsection{Scalar boundedness} \label{sec:scalar}
Keeping scalars strictly bounded in $Y \in [0,1]$ is important for chemical reactions~\cite{suh2025tvd}.
In the same spirit, we constrain $\vec M^\scB$ by directly implementing a bounded numerical scheme~\cite{larrouturou1991preserve} for the scalar flux, where the scalars $T$ and $Y$ are upwinded based on the nominal NN momentum $\bff^\rmc$, which is designed to be solenoidal~\cref{eqn:closure-chem} to conserve mass.
The non-conservative $\vec M^\scC$ is scaled by the $C_s = \dotomega$ prefactor, which limits the model to the reacting zone and keeps any violation near $c=0$ or $c=1$ as small as $\dotomega$.

\subsubsection{Equivariance} \label{sec:equivariance}
The governing equations~\cref{eqn:goveq-continuous} are frame-invariant~\cite{oberlack1997invariant}, but this is only preserved if $\vec M$ is \emph{equivariant}---to be defined below---under translation, rotation, and reflection in $\bfx$, uniform motion, and a shift in $t$~\cite{levy1971galilei}.
This benefits extrapolation: without it, for example, a model trained for a flame propagating in $x_3$ might yield different (and presumably inferior) results for a flame propagating in $x_1$.
Hence, non-invariance is a type of overfitting, which we avoid by building equivariance into $\vec M$.

\begin{figure}
	\centering
	\iffigures
    \ifcnf
	\includegraphics[width=\linewidth]{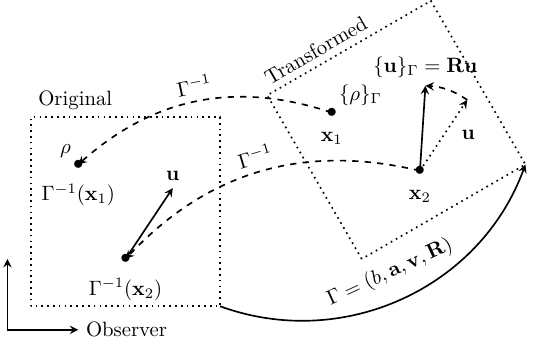}
    \else
	\includegraphics[width=0.5\linewidth]{figures/transformation.pdf}
    \fi
	\fi
    \caption{
        Representation of scalar and vector fields in a transformed frame of reference.
    }
\end{figure}

We first define a transformation $\Gamma \equiv (b, \bf a, \bf v, \bf R): (\bfx, t) \to (\bfx', t') = (\bf R \bfx + \bf v t + \bf a, t + b)$ in the Galilean group Gal(3) for an inertial frame of reference, parameterized by time shift $b \in \mathbb R$, space shift $\bf a \in \mathbb R^3$, uniform translation with velocity $\bf v \in \mathbb R^3$, and the rotation matrix $\bf R \in \rmO(3)$~\cite{levy1971galilei}.
A field variable $\phi(\bfx, t)$ transformed by $\Gamma$ and observed in a fixed frame of reference is denoted by $\{\phi\}_\Gamma $.
The explicit form depends on the type of $\phi$, as listed in~\cref{table:transformation}.
\begin{table}
    \centering
    \begin{tabular}{c c c}
        \toprule
        Type of $\phi$ & $\{\phi\}_\Gamma(\bfx, t)$ & Example \\
        \hline
        Scalar & $\phi(\Gamma^{-1}(\bfx, t))$ & $\rho$ \\
        Vector & $\bf R \phi(\Gamma^{-1}(\bfx, t)) + \bf v$ & $\bf u$ \\
        Pseudovector & $\det(\bf R) \bf R \phi(\Gamma^{-1}(\bfx, t))$ & $\nabla \times \bf u$ \\
        Rank--2 tensor & $\bf R \phi(\Gamma^{-1}(\bfx, t)) \bf R^\top$ & $\nabla \bf u$ \\
        \bottomrule
    \end{tabular}
    \caption{
        Explicit forms of $\{\phi\}_\Gamma$ under transformation $\Gamma = (b, \bf a, \bf v, \bf R)$, equivalent to rigid body transformations.
    }
    \label{table:transformation}
\end{table}

The quantity $\phi$ is \emph{invariant} if and only if
\begin{equation} \label{eqn:def-invariance}
	\{\phi\}_\Gamma = \phi,
\end{equation}
and a map $f(\phi)$ is \emph{equivariant} under $\Gamma$ if and only if
\begin{equation} \label{eqn:def-equivariance}
	f(\{\phi\}_\Gamma) = \{f(\phi)\}_\Gamma.
\end{equation}
The question is whether or not $\vec M$ satisfies~\cref{eqn:def-equivariance}:
\begin{equation}
    \{ \vec M(\vec Q^\theta) \}_\Gamma \overset{?}{=} M(\{\vec Q^\theta\}_\Gamma).
\end{equation}
Equivariance of $\vec M$ requires equivariance of its composing NN models:
\begin{equation} \label{eqn:equivariance-nn}
	\{\vec \calN(\vec Q^\theta)\}_\Gamma = \vec \calN(\{\vec Q^\theta\}_\Gamma).
\end{equation}
The model input~\cref{eqn:input} is equivariant for time and space shifts $(b \neq 0, \bf a \neq \bf 0)$, relative motion $(\bf v \neq \bf 0)$, and reflection about the origin $(\bf R = -\bf I)$.
Hence, full equivariance of $\vec \calN$ only requires additional rotational equivariance for $\bf R \neq \pm \bf I$.

As for the governing equations, this will not be achieved precisely for a mesh discretization.
However, enforcing a discrete equivalent should improve adherence.
Our square-mesh discretization is isotropic for 24 feasible orientations.
Therefore, these orientations are obvious choices to enforce equivariance of $\vec \calN$.
The averaging $\langle \; \rangle_G$ introduced in \cref{eqn:closure-turb,eqn:closure-chem} duplicates the input $\vec Q^\theta$ for all transformations in a finite subgroup $G$ of the orthogonal group $\rmO(3)$, evaluates the model on each branch, and averages the outputs from all branches:
\begin{equation} \label{eqn:G-average}
	\left\langle \vec \calN(\vec Q^\theta) \right\rangle_G \equiv \frac{1}{\vert G \vert} \sum_{\Gamma \in G} \left\{ \vec \calN(\{ \vec Q^\theta \}_\Gamma) \right\}_{\Gamma^{-1}},
\end{equation}
where $\vert G \vert$ is the size of $G$, so that the result is the average over $G$.
Then, the group-averaged $\vec \calN$ is equivariant under any $\Gamma' \in G$: in other words, transforming the input with $\Gamma'$ is equivalent to transforming the output with $\Gamma'$.
This is equivalent to other equivariant ML representations for finite transformation groups~\cite{cohen2016group,laptev2016ti}.

Of course, this adds cost.
If we consider all 24 orientations, computing~\cref{eqn:G-average} evaluates the same $\vec{\calN}$ model 24 times, so both the computing time and required memory increase by a factor of 24.
However, we will show in~\cref{sec:constraint-results} that even an SO(3) subgroup of $90^\circ$ rotation matrices with $\vert G \vert = 3$ provides significant improvement.

\subsubsection{Residual-consistency} \label{sec:apricon}
The last property is related to distinct regimes of this flow: the upstream turbulence, the flame, and the post-flame downstream.
It is expected that $\vec R$ in~\cref{eqn:goveq-les-ml} requires different augmentation depending on the locally active mechanisms. 
If $\vec R$ is sufficiently trusted in some region, it does not need a closure there.

\begin{figure*}
	\centering
	\iffigures
	\includegraphics[width=\linewidth]{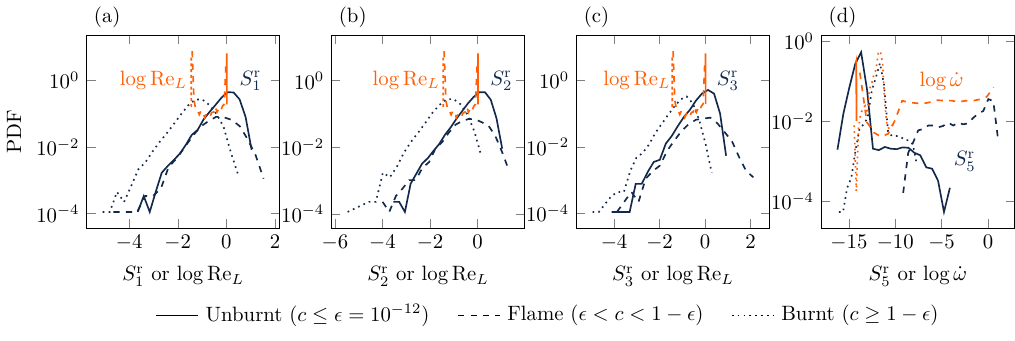}
	\fi
    \caption{
        Proability density function (PDF) of magnitudes of the exact SGS residuals of (a) $\rho u_1$, (b) $\rho u_2$, and (c) $\rho u_3$-equations, and (d) $\rho Y$-equation with their proxies: $\log \Re_L$ for $S_1^\rmr$, $S_2^\rmr$, and $S_3^\rmr$ and $\log \dotomega$ for $S_5^\rmr$.
        PDFs are sampled from all mesh points of a single DNS field, conditioned with the flame progress variable $c=1-Y$ on each mesh point.
        Residual of the $\rho E$-equation, $S_4^\rmr$, is not shown since the residual of the $\rho E$-equation can be simply expressed as a combination of $S_i^\rmr$ ($i \in \{1,2,3\}$) for the kinetic energy and $S_5^\rmr$ for the scalar transport and the chemical reaction.
    }
	\label{fig:pdf-scale}
\end{figure*}

To examine this, we first define an exact SGS residual of the $i$-th component of $\vec R$,
\begin{equation} \label{eqn:residual}
	R_i^\rmr \equiv \filter R_i(\vec q; \DelDNS) - R_i(\filter \vec q; \DelLES),
\end{equation}
and use it to define a local strength scale in~\cref{fig:pdf-scale}:
\begin{equation} \label{eqn:scale-residual}
	S_i^\rmr \equiv \log \vert R_i^\rmr \vert.
\end{equation}
Between the upstream higher-Re and downstream lower-Re regions, we expect $\bbE[S_i^\rmr \vert c=0] > \bbE[S_i^\rmr \vert c=1]$ for the momentum equations~\cref{eqn:goveq-continuous} for $i \in \{1,2,3\}$, which~\cref{fig:pdf-scale} supports.
We can similarly anticipate that the $\rho Y$-residual $R_5^\rmr$ is negligible away from the flame: $\bbE[S_5^\rmr \vert c \in (0,1)] \approx 0$ and $\bbE[S_5^\rmr \vert c \not\in (0,1)] \to -\infty$.

Closure models should be consistent with these or similar regime designations.
Most simply, the closure model should scale as the residual of the governing equations that represents the missing physics:
\begin{equation} \label{eqn:order-consistency}
	\log \vert M_i(\vec Q^\theta) \vert \propto \bbE[S_i^\rmr \vert \vec Q^\theta].
\end{equation}
This is a nominal `residual-consistency' that ensures that $M_i$ vanishes as $\Re \to 0$ for $i \in \{1,2,3\}$ and as $c \to 0$ or $c \to 1$ for $i=5$.

Enforcing~\cref{eqn:order-consistency} should be considered a design choice to keep $\vec M$ consistent with the represented physics: any nonzero $\vec M$ is simply wrong if the represented physics is perfectly trusted.
It also facilitates learning, as a model limited to correcting only the error of $\vec R$ has less to learn if the entire $\vec R$ were targeted~\cite{dresdner2022learning}.
Most importantly, it aligns with the main premise of closure in that the model should \emph{complement} the known physics.

Since our model $\vec M$ in~\cref{eqn:mlp} is trained to match the LES solution $\vec Q^\theta$, not the residual $\vec R^\rmr$, an explicit rescaling aids consistency between $\vec M$ and $\vec R^\rmr$.
Except for $\vec M^\scB$ and its $\bff^\rmc$ in~\cref{eqn:closure}, which automatically complies with~\cref{eqn:order-consistency} by design, we scale $\btau^\rmc$ and $s^\rmc$ with $C_\tau = \Re_L$, where $\Re_L = UL/\nu$ is the local Reynolds number, and $C_s = \dotomega$ as~\cref{eqn:closure-chem}.
This choice of prefactor is supported by the resemblance of $\log\Re_L$ and $\log\dotomega$ to $\bbE[S_i^\rmr \vert \vec Q^\theta]$ per regime in~\cref{fig:pdf-scale}.
Of course, the chosen scaling factors are specific for our demonstration, where a single variable $c$ sufficiently distinguishes different regimes, and the $T$-dependent $\Re_L$ and $\dotomega$ coincide with $\bbE[S_i^\rmr]$ for $i \in \{1,2,3\}$ and $\bbE[S_5^\rmr]$, respectively.
Other factors might be equally effective for scaling if available, such as $T^{-1}$ instead of $\Re_L$ for its effect on viscosity, or $Y(1-Y)$ instead of $\dotomega$, with the latter analogous to an eddy-break-up model~\cite{spalding1977development} or a flame surface density model~\cite{hawkes2000flame}.
Their primary purpose is to promote the residual-consistency by focusing training on regions with inaccurate dynamics.

\section{Model training} \label{sec:optimization}
We start with our embedded training approach in~\cref{sec:embedded} and then compare it with the unembedded approach in~\cref{sec:unembedded}.

\subsection{Embedded training} \label{sec:embedded}
The goal is to match the statistical QoIs from~\cref{sec:challenge}: $E(\kappa_\perp)$ for the turbulence and $s_\rmf$ and $A_\rmf$ for the flame dynamics.
However, matching the average directly entails a long time integration, which results in the Lyapunov growth of gradients.
To avoid this, we match the solution $\vec Q^\theta$ to the trusted $\filter \vec q$ for just $u_1$ and $Y$ over many short time segments: $N_{t,\opt} = 20$ time steps or $\tau_\opt = N_{t,\opt} \Delta t_\LES = 0.01$.
It will be shown in~\cref{sec:constraint-results} that the learned correction for $u_1$ indeed equally contributes to $u_2$ and $u_3$ in application, while a non-equivariant model fails to do so.

The specific training loss is $J_{\vec Q} \equiv J_u + J_Y$, with
\begin{equation} \label{eqn:loss-u}
	J_u = \frac{1}{2} \left\langle [u_1^\theta(\bfx, t_0 + \tau_\opt) - u_1(\bfx, t_0 + \tau_\opt)]^2 \right\rangle_\bfx,
\end{equation}
and
\begin{equation} \label{eqn:loss-c}
	J_Y = \frac{1}{2} \left\langle [Y^\theta(\bfx, t_0 + \tau_\opt) - Y(\bfx, t_0 + \tau_\opt)]^2 \right\rangle_\bfx,
\end{equation}
where each $(u_i^\theta, Y^\theta)$ and $(u_i, Y)$ are based on $\vec Q^\theta$ and $\filter \vec q$, respectively, and we initialize $\vec Q^\theta = \filter \vec q$ at each $t_0$.
With the adjoint-computed sensitivity, parameters in $\vtheta$ are then updated with the Adam optimizer~\cite{kingma2014adam} with a default learning rate of $10^{-3}$.
The output layer units are initialized to zero to avoid strong transients, and the rest are Xavier-initialized.
Mismatches~\cref{eqn:loss-u,eqn:loss-c} are averaged over 32 different $t_0$ instances forming a minibatch size of 32, randomly sampled every iteration from the set of 401 DNS fields.
Training progress is shown in~\cref{fig:loss}(a).

\subsection{For comparison: unembedded approach} \label{sec:unembedded}
\begin{figure}
	\centering
	\iffigures
    \ifcnf
	\includegraphics[width=0.8\linewidth]{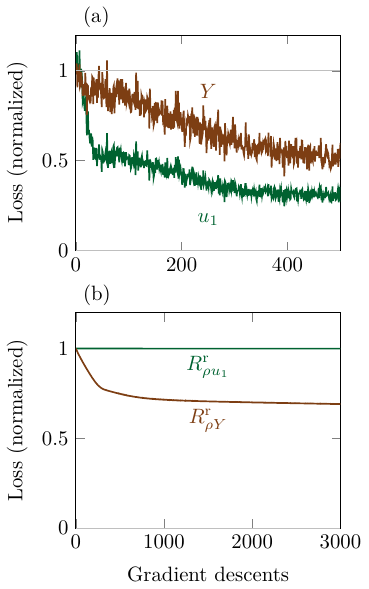}
    \else
	\includegraphics[width=0.8\linewidth]{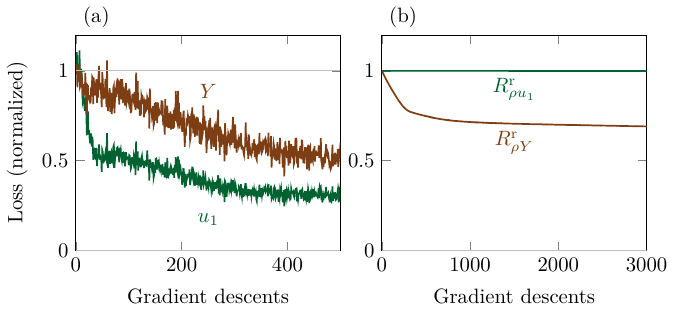}
	\fi
	\caption{
        Learning curves for (a) embedded and (b) unembedded training. Mismatches are normalized by their initial values.
    }
	\label{fig:loss}
\end{figure}

For comparison, the same NN models $\vec \calN$ in~\cref{eqn:closure} are also trained to directly match the residual $\vec R^\rmr$ with respect to the DNS,
\begin{equation} \label{eqn:loss-rhs}
	J_{\vec R} \equiv \frac{1}{2} \left\langle \Vert \vec M(\filter \vec q(\bfx, t_0)) - \vec R^\rmr(\vec q(\bfx, t_0)) \Vert_2^2 \right\rangle_\bfx,
\end{equation}
averaged over the 401 saved fields covering $t_0 \in [0,4]$.
This unembedded training is far simpler in that it only matches the closure target, but that data are rarely available in trusted form, and the governing equations do not constrain the process.
Parameters $\vtheta$ are initialized and updated with the same settings as in~\cref{sec:embedded}.
\Cref{fig:loss}(b) shows that the optimized model corrects $R^\rmr_{\rho Y}$ but barely changes $R^\rmr_{\rho u_1}$ mismatch.
With this training regimen, the model simply cannot match the mismatch data, which is essentially stochastic, as further discussed in~\cref{sec:irreducibility}.

\section{LES results} \label{sec:les}
All LES cases in this section run from the same initial condition $\filter \vec q$ at $t=0$ until $t=4$, so they extrapolate beyond the training $\tau_\opt = 0.01$.

\subsection{Predictions} \label{sec:prediction}
\begin{figure*}
    \iffigures
    \centering
    \includegraphics[width=\linewidth]{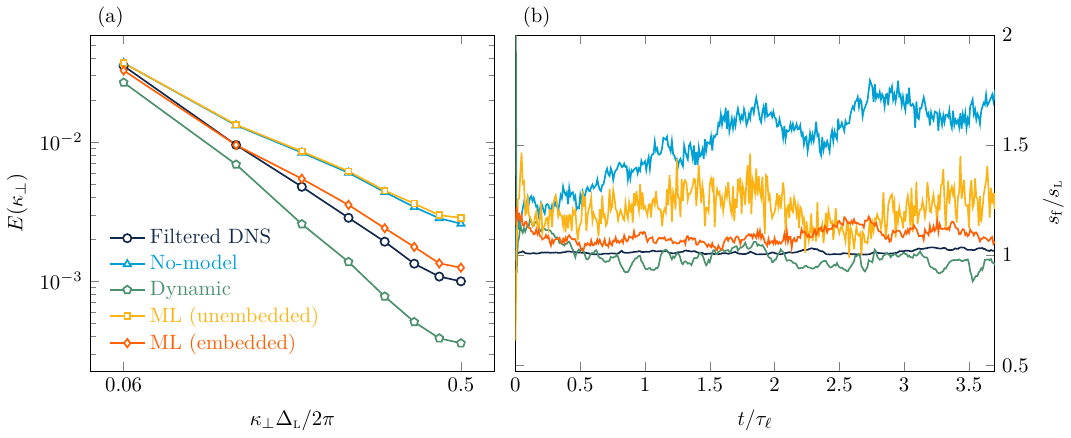}
    \fi
    \caption{
        (a) One-dimensional spanwise energy spectra and (b) predicted local flame speed $s_\rmf$ versus time.
    }
    \label{fig:results-turb+flame}
\end{figure*}

The models are compared in~\cref{fig:results-turb+flame}, along with a standard model that closes the SGS transport with the dynamic procedure~\cite{moin1991dynamic}, using the same second-order centered finite-difference scheme.
For this case, the SGS chemical reaction is closed with a Beta distribution for the SGS progress variable, with the SGS variance of the progress variable dynamically modeled as well~\cite{pierce1998dynamic}.

\begin{figure*}
    \iffigures
	\centering
    \includegraphics[width=\linewidth]{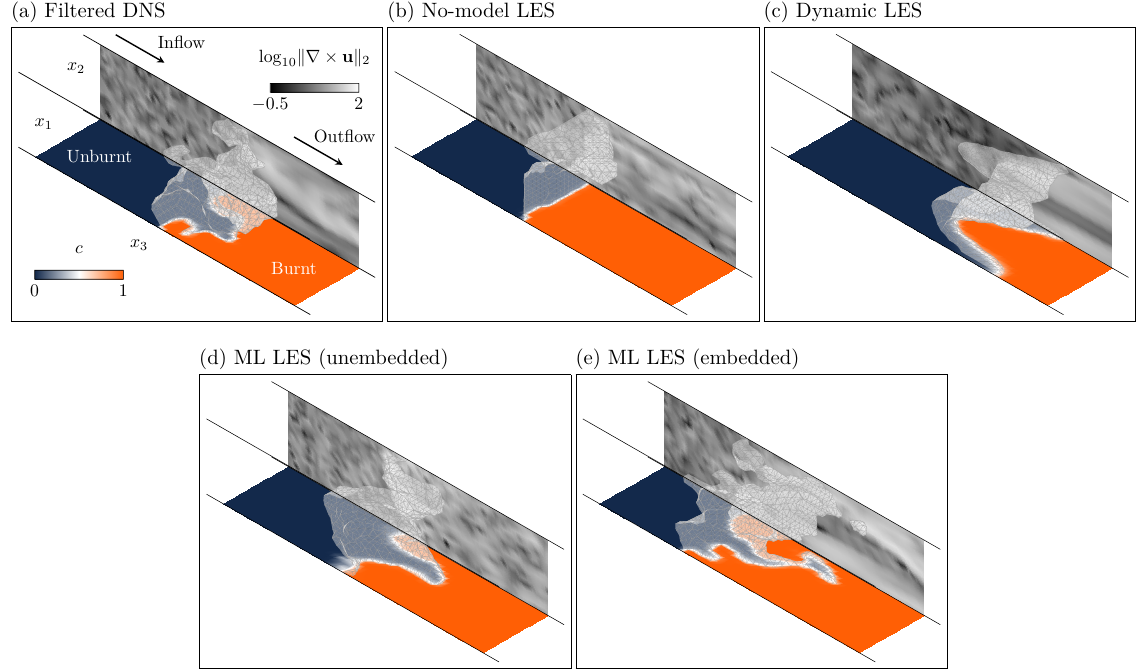}
    \fi
    \caption{
        Visualizations of flow fields and flames from simulations: (a) filtered DNS at $t=0$, LES (b) without an explicit closure, with (c) dynamic model, (d) unembedded ML closure, and (e) embedded ML closure.
        All LES cases are visualized at $t=3.7$.
        Note that the visualized filtered DNS case is the initial condition for all LES cases.
        Flame surfaces show $Y=0.18$.
        Vorticity magnitude and the flame progress variable $c=1-Y$ are plotted in two-dimensional planes.
    }
	\label{fig:flame3d}
\end{figure*}

\Cref{fig:results-turb+flame} shows that only the embedded ML model well predicts both the energy spectrum and the local flame speed.
The unembedded ML model barely affects the turbulence spectrum, although it does improve the flame speed prediction compared to the no-model LES, consistent with its learning curves in~\cref{fig:loss}(b).
The trained models also predict qualitatively correct flame distortions (\cref{fig:flame3d}).
The dynamic model fails to predict the turbulence as anticipated for the second-order finite-difference scheme with such coarse $\DelLES > \bar\Delta$~\cite{kravchenko1997effect}, though it does well predict the flame speed.
Good agreement for $s_\rmf$ is possible, despite their poor predictions for $E(\kappa_\perp)$, because only the large turbulent motions affect the flame dynamics in this regime.

Overall, only the embedded ML model was able to predict both turbulence and flame accurately.
The unembedded ML model and the dynamic model did not account for the coupling.

\subsection{Effectiveness of enforced equivariance} \label{sec:constraint-results}
\begin{figure*}
    \iffigures
	\centering
	\includegraphics[width=0.8\linewidth]{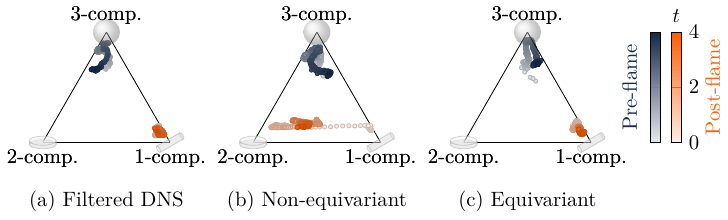}
    \fi
	\caption{
        Barycentric map of the Reynolds stress anisotropy tensor plotted for (a) filtered DNS, (b) LES with the non-equivariant model, and (c) LES with the equivariant ML model as the flow evolves from a downsampled DNS initial condition at $t=0$.
    }
	\label{fig:barycentric}
\end{figure*}

Following a standard procedure~\cite{banerjee2007presentation}, the reduced eigenvalues of the Reynolds stress anisotropy tensors for the filtered DNS, LES with the equivariant model, and LES with the non-equivariant model are plotted on barycentric maps in~\cref{fig:barycentric}.
The non-equivariant model simply omits $\langle \; \rangle_G$ in~\cref{eqn:closure-turb,eqn:closure-chem}.
The Reynolds stress is computed for each time instance both upstream $(c<0.01)$ and downstream $(c>0.99)$,
\begin{equation}
	\overline{\phi}(t,c) = \langle \phi(\bfx, t) \, \vert \, c^\circ(\bfx, t) = c \rangle_\bfx.
\end{equation}

While the nominally isotropic pre-flame turbulence remains close to the three-component limit, preserving near-isotropy in all cases shown in~\cref{fig:barycentric}, they differ downstream of the flame.
The equivariant trained LES shows flow close to the correct one-component limit of the DNS as the thermal expansion stretches the flow.
However, the prediction from the non-equivariant LES is biased toward the two-component limit.
It simply does not correct post-flame $u_2$ and $u_3$.

\section{Discussion} \label{sec:discussion}

\subsection{Cost}

\begin{table}
	\centering
	\begin{tabular}{c c c}
		\toprule
		& Prediction & Training \\
		\hline
		DNS & 14.9 & \\
		No-model LES & 0.093 & \\
		ML LES & 0.106 & 7.5 \\
		\bottomrule
	\end{tabular}
	\caption{
        Wall times (h) for prediction up to $t=1$ and adjoint-based training (only for ML LES) for the $J_{\vec Q}$-optimization over $\tau_\opt$ for 500 iterations, each with 20 time steps for both forward and adjoint solves.
        All cases for prediction ran on a single AMD MI300A GPU.
        Training used 32 GPU ranks for concurrent forward and adjoint solves for 32 different initial conditions.
    }
	\label{table:cost}
\end{table}
Wall times for both training the model and predicting are summarized in~\cref{table:cost}.
Predictions were run on a single dedicated AMD MI300A GPU.
Training includes forward and adjoint solves for 32 different initial conditions, all run concurrently on a total of 32 GPU ranks.
For the LES prediction, the trained ML model requires 14\% more time than the no-model LES, while maintaining a huge speedup over the DNS.
Adjoint-based training includes 20 forward PDE solves and 20 adjoint PDE solves per gradient descent iteration, with the adjoint solves taking most of the computation time.
The total training time for 500 iterations for $J_{\vec Q}$ is 7.5 hours.

\subsection{Benefits of embedded training} \label{sec:irreducibility}

\subsubsection{Irreducibility}
A perfect match with $J=0$ is not anticipated because the model underfits and only converges to its optimal estimate of the target~\cite{langford1999optimal,moreau2006optimal,vollant2017subgrid}.
In essence, the model inputs cannot explain all the given data, as analyzed for turbulence by Langford and Moser~\cite{langford1999optimal}.
Part of it is effectively aleatoric and so irreducible.
This sets a lower bound on $J$.

To explain the limits seen in~\cref{fig:loss}(b), we quantify the irreducibility in the training loss for the unembedded training approach for~\cref{eqn:loss-rhs}.
To do this, we first express the loss as
\begin{equation} \label{eqn:loss-rhs-stat}
	\langle J_{\vec R} \rangle = \Big\langle \big\Vert \vec M(\filter \vec q) - \vec R^\rmr(\vec q) \big\Vert_2^2 \Big\rangle_{\vec q},
\end{equation}
the average $L^2$-distance between the closure and the exact residual for all $\vec q(\bfx, t_0)$.
Based on the law of total variance~\cite{moreau2006optimal}, a lower bound of~\cref{eqn:loss-rhs-stat} is 
\begin{equation}\label{eqn:loss-rhs-bound}
    \langle J_{\vec R} \rangle \geq \Big\langle \big\Vert \vec M_\mathrm{opt}(\filter \vec q) - \vec R^\rmr(\vec q) \big\Vert_2^2 \Big\rangle_{\vec q},
\end{equation}
where $\vec M_\mathrm{opt}$ is the \emph{optimal estimate} for the target $\vec R^\rmr$.
That is the $\vec M$ that would minimize $\langle J_{\vec R} \rangle$ given $\filter \vec q$:
\begin{equation} \label{eqn:optimal-estimate}
    \vec M_\mathrm{opt}(\filter \vec q) \equiv \big\langle \vec R^\rmr(\vec q^{\,\circ}) \, \vert \, \filter \vec q^{\,\circ} = \filter \vec q \big\rangle_{\vec q^{\,\circ}},
\end{equation}
which is the expectation of $\vec R^\rmr(\vec q^{\,\circ})$ over all possible $\vec q^{\,\circ}$ (in a hypothetical DNS solution space) that form the same input for $\vec M$ as $\vec q$ does, equivalent to the ``ideal LES'' evolution by Langford and Moser~\cite{langford1999optimal}.

The nonzero lower bound~\cref{eqn:loss-rhs-bound} is irreducible because of the information loss caused by the $\filter$ operation and the locality of the input for $\vec M$~\cite{lozano2022information}.
In short, the availability of $\filter \vec q$ is insufficient to fully recover $\vec q$ and thereby the $\vec R^\rmr(\vec q)$ that would exactly match.
There exist $\vec q$ and $\vec q^{\,\circ}$ that form the same inputs $\filter \vec q^{\,\circ} = \filter \vec q$ but would correspond to different targets $\vec R^\rmr(\vec q^{\,\circ}) \neq \vec R^\rmr(\vec q)$.

The irreducible limit~\cref{eqn:loss-rhs-bound} in the training loss~\cref{eqn:loss-rhs} can be estimated and compared with the empirical lower bound observed from~\cref{fig:loss}(b).
We demonstrate this for the $R^\rmr_{\rho Y}$ mismatch.
This requires its conditional average binned with $\filter \vec q$, which we approximate using $\filter c$ as the conditioning variable as one representative component of $\filter \vec q$.
The optimal estimate for $R^\rmr_{\rho Y}$ is then
\begin{equation}
    M_{\rho Y,\opt}(\filter \vec q) = \langle R^\rmr_{\rho Y}(\vec q^{\,\circ}) \, \vert \, \filter c^\circ = \filter c \rangle_{c^\circ}
\end{equation}
for $\filter c$ and $c^\circ$ derived from $\filter \vec q$ and $\vec q^{\,\circ}$, respectively, and the anticipated lower bound of the $R^\rmr_{\rho Y}$ mismatch is
\begin{equation}
    \label{eqn:loss-rhoY-bound}
	\langle J_{R_{\rho Y}} \rangle \geq \Big\langle \big[ M_{\rho Y,\opt}(\filter \vec q) - R^\rmr_{\rho Y}(\vec q) \big]^2 \Big\rangle_{\vec q},
\end{equation}
which is the mean conditional variance of $R^\rmr_{\rho Y}$ given $\filter c$.
\begin{figure}
    \iffigures
	\centering
    \ifcnf
	\includegraphics[width=0.8\linewidth]{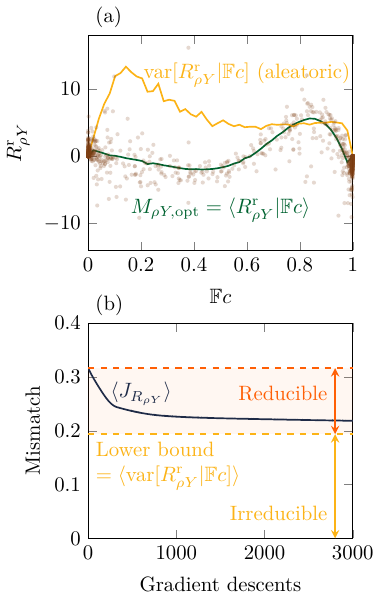}
    \else
	\includegraphics[width=0.8\linewidth]{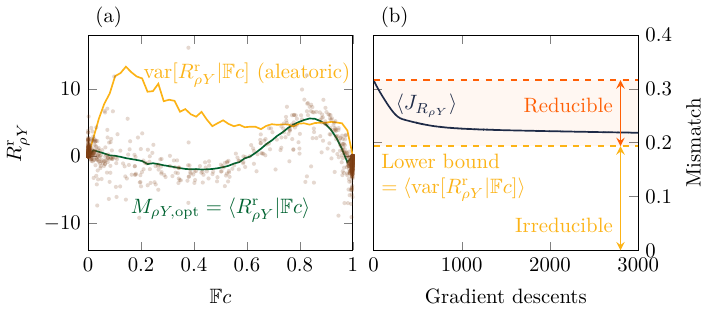}
    \fi
    \caption{
        (a) Mean and variance of the example right-hand side $R^\rmr_{\rho Y}$ conditioned by $\filter c$ as the input variable, computed from the DNS.
        The scatter plot shows the actual data $(\filter c, R^\rmr_{\rho Y})$.
        (b) Learning curve for the $J_{R_{\rho Y}}$ with the decomposition.
    }
	\label{fig:loss-explained}
\end{figure}

\Cref{fig:loss-explained}(a) shows the decomposition of $R^\rmr_{\rho Y}$ versus the conditioning variable $\filter c$.
The conditional mean is the optimal estimate, and the conditional variance is the irreducible aleatoric uncertainty in $R^\rmr_{\rho Y}$ for the model.
The theoretical lower bound of $\langle J_{R_{\rho Y}} \rangle$~---~the mean of $\mathrm{var}[R^\rmr_{\rho Y} \vert \filter c]$~\cref{eqn:loss-rhoY-bound}~---~is plotted with the learning curves from the unembedded training in~\cref{fig:loss-explained}(b).
Its agreement with the empirical lower bound of $\langle J_{R_{\rho Y}} \rangle$ in~\cref{fig:loss-explained}(b) suggests that most of the reducible error in the mismatch is indeed corrected by the trained model, so the model is capable of learning the effective dynamics from the training data.
Subsequently, the negligibly reduced training loss in~\cref{fig:loss}(b) indicates that irreducibility dominates the $R^\rmr_{\rho u_1}$ mismatch.
Even if a reducible signal exists in~\cref{eqn:loss-rhs}, such as the obvious lack of SGS dissipation in the no-model LES, it is masked so the model is unable to learn it.

\subsubsection{Amplified reducibility in embedded training}

\begin{figure*}
    \iffigures
	\centering
    \ifcnf
	\includegraphics[width=0.9\linewidth]{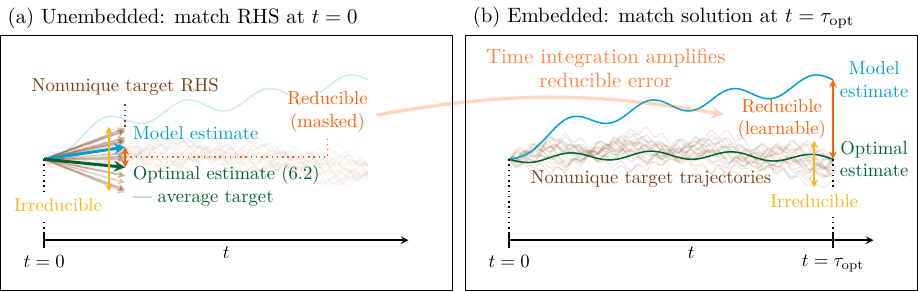}
    \else
	\includegraphics[width=0.9\linewidth]{figures/embed-vs-unembed.pdf}
    \fi
    \fi
    \caption{
        Reducible and irreducible errors in (a) unembedded and (b) embedded training.
        Plots are based on synthetic data for illustration.
    }
	\label{fig:irreducibility}
\end{figure*}

Both embedded and unembedded models are susceptible to irreducibility because predictions are the time integration of the governing equations, so they include the nonlinear accumulation of the residual of the governing equations~\cite{suh2026thesis}.
What matters is whether these models can learn any reducible signal from the training data that is relevant to predictions.

\Cref{fig:irreducibility} illustrates how irreducibility arises in unembedded and embedded approaches and how it amplifies over time integration for the embedded approach.
Both cases match LES to DNS: the unembedded approach matches the right-hand side (RHS) at $t=0$, while the embedded approach matches the solution over the finite $\tau_\opt$.
Targets are not unique in both cases, as the filtering and downsampling remove information from DNS, so infinitely many candidates of ``exact'' DNS exist for an LES~\cite{langford1999optimal}.
The trained model then only converges to the optimal estimate, the average of such nonunique targets, and the variance of the target quantities with respect to the optimal estimate indicates the irreducible aleatoric uncertainty in data.
Such irreducibility masks the reducible error in the unembedded approach, so the model is unable to infer any information from the training data.
This leads to unsuccessful learning as in~\cref{fig:loss}(b).

For the embedded approach, the small reducibility that was negligible at $t=0$ is amplified in time, appearing as a correctable signal in the solution mismatch that informs the models of the relevant dynamics.
A concrete example is the cumulative effect of erroneous viscous dissipation.
The lack of SGS dissipation itself has small reducibility.
The accumulated effect of this is large, resulting in an overly energetic flow.
The growing error in energy better informs the models of the needed SGS dissipation.

There is no guarantee that the time-integrated deviation of $\vec R^\theta$ only amplifies the reducible component, though it appears to do so for our demonstration.
In our case, the irreducibility in the solution mismatch is bounded because the DNS solution itself is statistically stationary.
In such a case, where the time-integration amplifies the reducibility with the bounded irreducibility, matching predictions guarantees outperformance of a model trained to match governing equations such as for PINNs~\cite{raissi2019physics}.

\section{Conclusion} \label{sec:conclusion}

We introduced an embedded training closing an LES of a turbulent flame.
Although the demonstration was for simple chemistry, the developed ML closure modeling framework lays the groundwork for future development of other data-driven closures for more complex scenarios, such as those with multi-species chemical reactions.

The constrained model provides essential features for closure models for other reacting flows.
One remaining question is whether the model may have been able to learn the constraints we identified as beneficial.
More generally, is it possible to build a model for the mixed desiderata reflecting both quantitative and qualitative soundness, without human intervention, such as our effort of enforcing constraints?
This is an optimization task with multiple objectives, where some objectives that represent constraints warrant higher priorities than the others.
Unless such a hierarchy of priorities of tasks is reflected in the training loss or the optimization procedure, the trained model likely violates overlooked constraints, either because of the conflicting objectives or because ML tends to prioritize optimizing for an easier task~\cite{suh2025tvd}.
We believe that known constraints are best enforced in the model formulation and training.

Importantly, embedded training amplifies reducible signals in data, as seen in~\cref{sec:irreducibility}, which supports a general superiority over many other approaches that aim to learn from governing equations~\cite{wang2018investigations,beck2019deep,park2021toward,kang2023neural,lozano2023machine,benjamin2024neural}, such as by minimizing~\cref{eqn:loss-rhs}.
Of course, this advantage needs to be balanced against the challenges, including the exploding gradients for chaotic systems and the cost of adjoint-based training~\cite{dehtyriov2025orans,hickling2026ogf}.

\ifcnf

\section*{CrediT authorship contribution statement}
Seung Won Suh: Conceptualization, Methodology, Software, Validation, Formal analysis, Investigation, Resources, Data curation, Writing - original draft, Visualization.
Jonathan F. MacArt: Conceptualization, Methodology, Software, Writing - review \& editing, Supervision.
Luke N. Olson: Conceptualization, Methodology, Writing - review \& editing, Supervision, Project administration, Funding acquisition.
Jonathan B. Freund: Conceptualization, Methodology, Writing - review \& editing, Supervision, Project administration, Funding acquisition.

\section*{Declaration of competing interest}
The authors declare that they have no known competing financial
interests or personal relationships that could have appeared to
influence the work reported in this paper.

\fi

\section*{Acknowledgments}
This material is based in part upon work supported by the Department of Energy, National Nuclear Security Administration, under Award Number DE-NA0003963.

\appendix

\ifcnf
\bibliographystyle{cnf-num}
\else
\bibliographystyle{ieeetr}
\fi

\bibliography{refs}

\end{document}